# TITLE PAGE

A clinically relevant online patient QA solution with daily CT scans and EPID-based in vivo dosimetry: A feasible study on rectal cancer

Liyuan Chen[*,1,2,3], Zhiyuan Zhang[*,1,2,3], Lei Yu[1,2,3], Jiyou Peng[1,2,3], Bin Feng[1,2,3], Jun Zhao[1,2,3], Yanfang Liu[4], Fan Xia[1,2,3], Zhang Zhen[1,2,3], Weigang Hu[#,1,2,3], Jiazhou Wang[#,1,2,3]

[1]Department of Radiation Oncology, Fudan University Shanghai Cancer Center, Shanghai 200032, China

[2]Department of Oncology, Shanghai Medical College, Fudan University, Shanghai 200032, China

[3]Shanghai Key Laboratory of Radiation Oncology, Shanghai 200032, China

[4]Radiation Therapy BU, United Imaging Healthcare, Shanghai 201807, China

* contribution equally

\# corresponding author

Acknowledgments: This work is supported by the Shanghai Committee of Science and Technology Fund (19DZ1930902, 21Y21900200), Xuhui District Artificial Intelligence Medical Hospital Cooperation Project (2021-012) and National Natural Science Foundation of China, Grant/Award Numbers: 11675042.

Corresponding author: JiaZhou Wang

Email: wjiazhou@gmail.com

Fax number: 86-21-6417 4774

Phone number: 86-18017312765

Corresponding author: Weigang Hu

Email: jackhuwg@gmail.com

Fax number: 86-21-6417 4774

Phone number: 86-15921292224

Running title: An online patient QA solution

# A clinically relevant online patient QA solution with daily CT scans and EPID-based in vivo dosimetry: A feasible study on rectal cancer

**Abstract**

Purpose: Adaptive radiation therapy (ART) could protect organs at risk (OARs) while maintain high dose coverage to targets. However, there still lack efficient online patient quality assurance (QA) methods, and it has been an obstacle to large-scale adoption of ART. We aim to develop a clinically relevant online patient QA solution for ART using daily CT scans and EPID-based in vivo dosimetry.

Materials and Methods: Ten patients with rectal cancer at our center were included. Patients' daily CT scans and portal images were collected to generate reconstructed 3D dose distributions. Contours of targets and OARs were recontoured on these daily CT scans by a clinician or an auto-segmentation algorithm, then dose-volume indices were calculated, and the percent deviation of these indices to their original plans were determined. This deviation was regarded as the metric for clinically relevant patient QA. The tolerance level was obtained using a 95% interval of the QA metric distribution. These deviations could be further divided into anatomically relevant or delivery relevant indicators for error source analysis. Finally, our QA solution was validated on an additional six clinical patients.

Results: In rectal cancer, the lower and upper tolerance of the QA metric for PTV $\Delta D95$ (%) were [-3.11%, 2.35%], and for PTV $\Delta D2$ (%) were [-0.78%, 3.23%]. In validation, the 68% for PTV $\Delta D95$ (%), and the 79% for PTV $\Delta D2$ (%) of the 28 fractions are within tolerances of the QA metrics. By using four or more out-of-tolerance QA metrics as an action level, there were 5 fractions (18%) have four or more out-of-tolerance QA metrics in validation patient dataset.

Conclusion: The online patient QA solution using daily CT scans and EPID-based in vivo dosimetry is clinically feasible. Source of error analysis has the potential for distinguishing sources of error and guiding ART for future treatments.

Key words: Quality assurance, In vivo dosimetry, Dose volume histograms, Adaptive radiation therapy

1. **Introduction**

Radiotherapy aims to deliver a high therapeutic dose to a tumor while minimizing exposure to the surrounding healthy tissue.[1] A typical radiotherapy treatment procedure includes the following steps: CT scanning, target and organ at risk (OAR) delineation, treatment planning, plan quality assurance (QA) and treatment delivery. This procedure assumes that the patient's anatomy is static during the whole treatment. This hypothesis may be inappropriate for many tumor sites, such as the rectum, and patients may experience anatomic changes during the whole treatment, such as tumor shrinkage, weight loss and bladder filling variations. These changes may affect the outcome of the treatment.[2] To consider these variations, the concept of adaptive radiation therapy (ART) was developed.[3-5]

ART, which was proposed by Yan et al. in 1997,[6] utilizes various feedback, such as daily images, to rapidly modify a radiotherapy plan during the whole treatment course. However, due to technical and resource limitations, the large-scale clinical practice of ART is still lacking. At technical level, there are three main obstacles in ART: The lack of high-quality image acquisition, rapid treatment planning and online plan QA solutions. With the emergence of MR-Linac,[7] CT-Linac[8] and artificial intelligence (AI) technology,[9] there may be solutions to the first two obstacles.[10,11] For online plan QA, more research is needed to find an appropriate solution.[12]

A traditional pretreatment patient-specific plan QA, which is usually performed on phantoms or devices, can be difficult to implement in the case of the patient on the couch.[13] Therefore, many alternative methods have been developed.[14-17] Among these methods, electronic portal imaging device (EPID)-based in vivo dosimetry measurements have been adopted by many researchers and venders for its benefits, such as fast image acquisition, high resolution, independence and ability to achieve 3D dose verification.[18,19]

Many studies have used deep learning based or deformable registration based methods to generate synthetic CT for dose calculation.[20-22] These methods may introduce additional uncertainty in dosimetry.[23,24] Strict alignment of voxel information between CBCT and planning CT need to be guaranteed but it is difficult to achieve in practice[25] and inaccurate HUs or image artifacts are still exist in synthetic CT.[26] 4% deviation in some dose-volume indices or 5% decrease in gamma passing rate (GPR) can be observed, and this degree of deviation could not be accepted in QA procedure.[26-28] Some researches about in vivo dosimetry based on EPID and CBCT also reported a similar way to synthesize CT for dose calculation.[29-32] The evaluation metrics in these studies were GPR, dose-volume index, dose at one point or volumetric changes. And there is no consensus about the tolerance threshold.

Many issues about online patient QA remain unresolved. First, patient QA metrics are weakly related to the clinic.[33] Commonly used metrics, such as the GPR[34,35], cannot reflect the dose deviation on critical OARs that concern physicians.[36] Second, there is no consensus how to set an appropriate tolerance for different tumor sites.[35,37-39] Third, especially for ART, the cause of the dose deviation may not be easily distinguished.[40,41] Deviation in the measured values may result from variation in a patient's anatomy or inaccuracy in the delivery devices.[35]

Aiming at these issues, we propose a clinically relevant online patient QA solution for ART. With diagnostic low-dose daily CT scans on which dose distribution can be directly calculated and EPID-based in vivo dosimetry, the dose distribution that a patient received can be reconstructed as accurately as possible. By comparing this reconstructed dose distribution to the planning dose distribution, a clinically relevant QA tolerance is established. Meanwhile, sources of error are investigated by comprehensively analyzing various types of metrics.

The solution directly connects to the dose-volume indices that physicians and physicists are concerned about, and a tolerance setting method based on the treatment history was proposed. Moreover, source of error analysis could help physicists with follow-up actions. To demonstrate the proposed solution, the implementation and validation results of data from patients with rectal cancer are presented as an example in this study.

We briefly describe the device and related algorithm used in Section 2.1. Then, the proposed QA solution is introduced in Section 2.2. The implementation and validation results of data from patients with rectal cancer are demonstrated in Section 2.3., while the detailed results are presented in the results section.

## 2. Materials and methods

*2.1 Device and algorithm*

All treatments were delivered on a uRT-Linac 506c medical linear accelerator (United Imaging Healthcare, Shanghai, China), which has a diagnostic-quality 16-slice helical CT imager coaxially attached to the gantry of a C-arm linac and is equipped with an amorphous silicon EPID XRD1642 (Varex Imaging, UT, USA). The EPID yields a 40.96 × 40.96-cm$^2$ detection area. Additional details of the device description can be found in our previous studies.[8,42]

A vendor provided a Monte Carlo-based dose reconstruction algorithm, which was directly integrated into TPS. This algorithm only uses EPID and patient CT scan data to calculate the reconstructed dose distribution. A series of algorithm verifications was performed before this study, and the details and results are presented in the supplement.

*2.2 Proposed online QA solution*

Figure 1 shows a general workflow to establish the QA protocol and criteria for a specific tumor site. Inspired by the concept of confidence intervals in statistics, a similar idea was used to determine the tolerance of dose deviations. Therefore, a group of patients with the same tumor site were required to establish the dose deviation distribution.

These patients were treated with routine radiotherapy procedures, and two types of data were collected during treatment, including pretreatment low-dose daily CT scans, which was used to perform image-guided radiation therapy (IGRT), and a transmission portal image for each fraction. These data were used to calculate a series of dose-volume indices and perform the following analyses: 1. QA metrics calculation; 2. Tolerance settings; and 3. Source of error analysis.

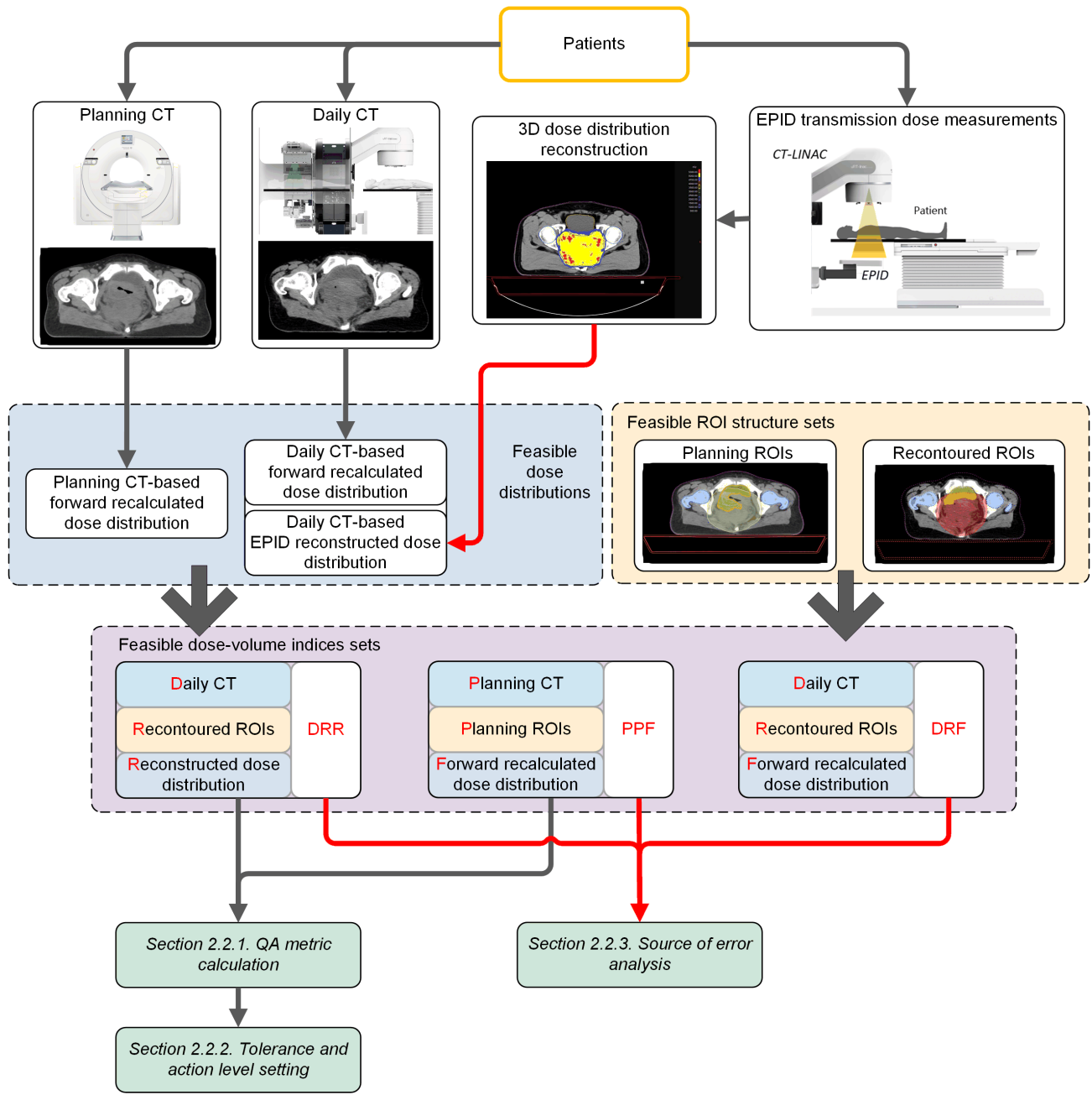

Figure 1. Workflow of established QA protocol. Abbreviations of the series of dose-volume indices used in this QA protocol are shown in red.

2.2.1 QA metric calculation

To estimate the actual dose that patients received in a single fraction, the dose distribution should be calculated by a dose reconstruction algorithm on daily CT scans with EPID data, and the dose-volume indices need to be calculated on recontoured region of interests (ROIs). The PTV of daily CT scans may need to be manually delineated by the same physician with the same protocols, while OARs can be delineated by an automatic segmentation algorithm if consistent with the treatment planning process.

The abbreviation DRR represents the dose-volume indices obtained from daily CT scans with recontoured ROIs and EPID-based reconstructed dose distribution. Compared to dose-volume indices from other combinations, the deviations between these specific indices and the original treatment planning indices are closest to the true deviations between treatment delivery and planning. Therefore, the percentage of this deviation was used as the QA metric of the dose-volume index.

*2.2.2 Tolerance and action level setting*

A critical issue for patient QA is the tolerance setting. Inspired by the concept of confidential intervals, a statistics-based tolerance level was proposed based on a series of dose-volume indices. These indices were selected based on clinical considerations, such as the D95 of the PTV.

Before dose-volume index calculations, physicians and physicists were required to review the daily CT scans and reconstructed dose distributions to ensure that there were no outliers in the sample data for the tolerance level setting. The 95% intervals are regarded as the tolerance range. The SW test was used to decide whether each index followed a normal distribution. A normal distribution was fitted if it was followed, and 95% interval was decided by [μ-2σ, μ+2σ], where μ is the mean value of the metric and σ is the standard deviation; otherwise, the ranges from the 2.5% to 97.5% quantiles were used as the 95% interval.

*2.2.3 Sources of error analysis*

The dose-volume deviation can be divided into anatomically relevant and delivery relevant deviations. By further analyzing the data of QA metric, the source of the error may be located. For a certain dose-volume index, the deviation between daily CT scans and planning CT scans with the same dose type (both are forward calculated or reconstructed) can be used to indicate the impact of anatomical variation, while deviation between reconstructed dose and forward calculated dose in the same CT scans can be used to indicate the impact of delivery errors. For example, (Equation 1), the deviation between DRR (daily CT scan with recontoured ROIs and reconstructed dose distribution) PTV D95 and planning (PPF, dose-volume indices obtained using planning CT scans with planning ROIs and forward calculated dose distribution) PTV D95 can be divided into two deviations: the deviation between DRR PTV D95 and DRF (daily CT scan with recontoured ROIs and forward calculated dose distribution) PTV D95 and the deviation between DRF PTV D95 and planning PTV D95. The former represents the delivery deviation, and the latter represents the impact of anatomical variation.

$$\frac{DRR\ PTV\ D95 - PPF\ PTV\ D95}{PPF\ PTV\ D95} = \frac{DRR\ PTV\ D95 - DRF\ PTV\ D95}{PPF\ PTV\ D95} + \frac{DRF\ PTV\ D95 - PPF\ PTV\ D95}{PPF\ PTV\ D95} \qquad (1)$$

### 2.3 Implementation and validation of rectal cancer

#### 2.3.1 Patients and data

Ten rectal cancer patients who were at our institution from March 2021 to July 2021 were enrolled in this study to establish the tolerance level. In total, 45 daily CT scans were taken from these patients. All patients were treated with 6 MV photon intensity-modulation radiation therapy (IMRT) or volume-modulation arc therapy (VMAT) with a prescribed dose of 50 Gy/25 fractions or 25 Gy/5 fractions.

#### 2.3.2 Dose-volume indices selection

Based on our clinical routine, 10 indices of the ROIs were selected in this study, including D95, D2, HI and CI of the PTV; D15 and D50 of the bladder; and D25 and D40 of the left or right femoral head (FH-L or FH-R). To demonstrate the clinical use of these indices, we set 4 QA metrics out of tolerance as action levels. For the patient who have more than 4 indices out of tolerance, a review by physicist will be required.

#### 2.3.3 Source of error analysis

The decomposition equations for rectal cancer used in this study are listed in Table 1.

Table 1. Deviation decomposition table

| Total deviation | Delivery relevant deviation | Anatomically relevant deviation |
|---|---|---|
| DRR PTV D95 - PPF PTV D95 | DRR PTV D95 - DRF PTV D95 | DRF PTV D95 - PPF PTV D95 |
| DRR PTV D2 - PPF PTV D2 | DRR PTV D2 - DRF PTV D2 | DRF PTV D2 - PPF PTV D2 |
| DRR Bladder D15 - PPF Bladder D15 | DRR Bladder D15 – DRF Bladder D15 | DRF Bladder D15 - PPF Bladder D15 |
| DRR Bladder D50 - PPF Bladder D50 | DRR Bladder D50 - DRF Bladder D50 | DRF Bladder D50 - PPF Bladder D50 |
| DRR FH-L D25 - PPF FH-L D25 | DRR FH-L D25 - DRF FH-L D25 | DRF FH-L D25 - PPF FH-L D25 |
| DRR FH-L D40 - PPF FH-L D40 | DRR FH-L D40 - DRF FH-L D40 | DRF FH-L D40 - PPF FH-L D40 |
| DRR FH-R D25 - PPF FH-R D25 | DRR FH-R D25 - DRF FH-R D25 | DRF FH-R D25 - PPF FH-R D25 |
| DRR FH-R D40 - PPF FH-R D40 | DRR FH-R D40 - DRF FH-R D40 | DRF FH-R D40 - PPF FH-R D40 |

Note: DRR, dose-volume indices obtained using daily CT scans with recontoured ROIs and reconstructed dose distribution; DRF, dose-volume indices obtained using daily CT scans with recontoured ROIs and forward calculated

dose distribution; and PPF, dose-volume indices obtained using planning CT scans with planning ROIs and forward calculated dose distribution. All deviations were divided by PPF dose-volume indices to obtain percent deviations.

Common anatomical variations were included in the clinical data of the 10 patients, while delivery errors were relatively small and not easy to distinguish. To demonstrate the effectiveness of the source of the error analysis, a few delivery errors were deliberately introduced into the original plan of the 10 patients. The error types and error magnitudes are shown in Table 2. Small errors within the tolerance of TG142[43] were not considered, and errors with too large a magnitude were almost impossible in the clinic and were also not included.

Table 2. Errors deliberately introduced to original plans

| Error type | Gantry angle (°) | Collimator angle (°) | MLC shift (mm) | MU scaling (%) |
|---|---|---|---|---|
| Magnitude | -3, -5, 3, 5 | -3, -5, 3, 5 | -3, -5, 3, 5 | 3, 5 |

*2.3.4 Solution validation*

To validate our proposed solution, an additional 6 rectal cancer patients with a total of 28 daily CT scans from April 2021 to November 2021 were enrolled in the study. These patients were treated with 6 MV photon IMRT or VMAT with a prescribed dose of 50 Gy/25 fractions, and their PTVs on daily CT scans were also recontoured by the same clinician so that the QA metric of them could also be evaluated.

3. **Result**

*3.1 QA metric and tolerance*

The distribution and the tolerance of the ten QA metrics are shown in Figure 2. All SW test results and specific values of tolerance ranges are shown in Table 3. For instance, the tolerance of PTV ΔD95 (%) is [-3.11%, 2.35%].

Table 3. SW test and tolerance of each QA metric

| Dose-volume index [%] | SW test p-value | Lower | Upper |
|---|---|---|---|
| PTV ΔD95 | 0.2355 * | -3.11% | 2.35% |
| PTV ΔD2 | 0.0129 | -0.78% | 3.23% |
| PTV ΔHI | 0.0002 | 3.84% | 145.81% |
| PTV ΔCI | 0.2047 * | -15.20% | -4.15% |
| Bladder ΔD15 | 0.0015 | -12.99% | 19.00% |

| | | | | |
|---|---|---|---|---|
| Bladder ΔD50 | 0.0332 | -20.98% | 28.31% | |
| FH-L ΔD25 | 0.0675 * | -2.25% | 8.16% | |
| FH-L ΔD40 | 0.1653 * | -2.85% | 10.93% | |
| FH-R ΔD25 | 0.8858 * | -3.56% | 5.09% | |
| FH-R ΔD40 | 0.1891 * | -2.59% | 7.38% | |

Note: * QA metrics with SW test p-value bigger than 0.05, normal distributions were fitted for them.

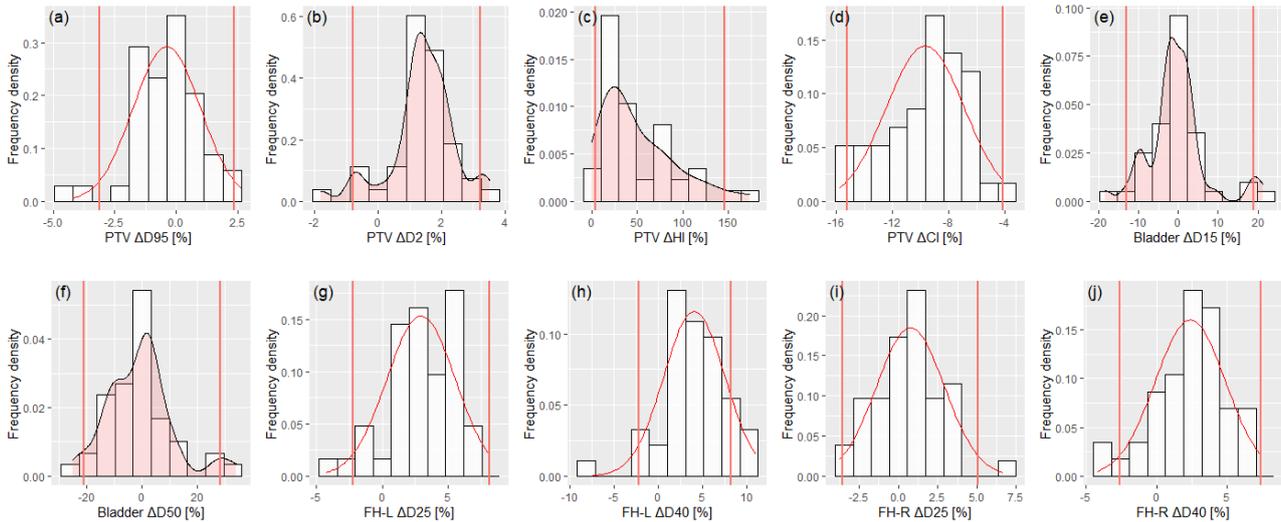

Figure 2. Distributions of the percent deviation of dose-volume indices. Red curves represent the distributions that satisfy a normal distribution, and these curves yield normal fitting results. The range between vertical red lines is the 95% interval of the percent deviation distribution.

3.2 Source of error analysis

Based on the data from one patient, the total deviation, anatomically relevant deviation and delivery relevant deviation during the whole treatment (5 fractions) are shown in Figure 3. In this patient's data, the main error relates to anatomical deviations. This deviation occurred mainly in the bladder, which means that this organ may have anatomical variation among treatments.

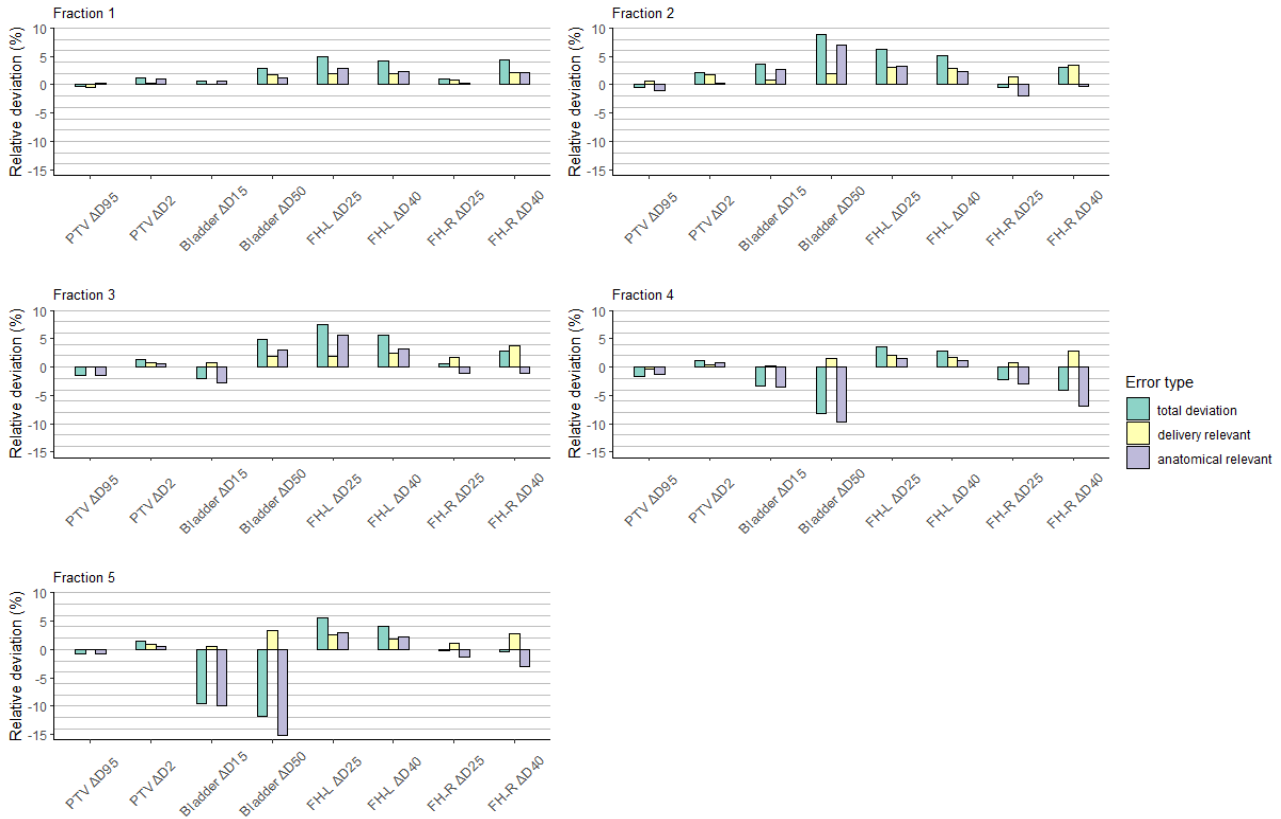

Figure 3. Error source analysis for the data from one patient. Green bars are the total percent deviation. Yellow and purple bars are delivery relevant and anatomically relevant deviations. The various deviations are defined in Table 2. For example, three deviations of PTV ΔD95 (%) are the total deviation: (DRR PTV D95 - PPF PTV D95)/ PPF PTV D95; delivery relevant: (DRR PTV D95 - DRF PTV D95)/ PPF PTV D95; and anatomically relevant: (DRF PTV D95 - PPF PTV D95)/ PPF PTV D95.

The tolerance calculated in Section 3.1 was used to evaluate these error-relevant deviations. Table 4 shows the number of dose-volume indices that are out of tolerance with or without delivery error introduced. For nonerror-introduced fractions, all delivery relevant deviation indices are within tolerance. There were no fractions with four or more out-of-tolerance indices about the three types of errors.

Table 4. The number of out-of-tolerance dose-volume indices with or without delivery error introduced

| | Number of out-of-tolerance indices for one fraction [indices] | Number of fractions | | |
|---|---|---|---|---|
| | | Anatomical + delivery [fractions] | Anatomical [fractions] | Delivery [fractions] |
| **Without delivery error introduced** | All indices within tolerance | 28 (62.2%) | 27 (60%) | 45 (100%) |
| | 1 index out of tolerance | 11 (24.4%) | 10 (22.2%) | 0 (0%) |
|  | 2 indices out of tolerance | 5 (11.1%) | 4 (8.9%) | 0 (0%) |
|  | 3 indices out of tolerance | 1 (2.2%) | 4 (8.9%) | 0 (0%) |
|  | Total | 45 (100%) | 45 (100%) | 45 (100%) |
|  | All indices within tolerance | 214 (34.0%) | 384 (61.0%) | 364 (57.8%) |
|  | 1 index out of tolerance | 175 (27.8%) | 151 (24.0%) | 114 (18.1%) |
|  | 2 indices out of tolerance | 116 (18.4%) | 62 (9.8%) | 72 (11.4%) |
| **Delivery error introduced** | 3 indices out of tolerance | 63 (10.0%) | 30 (4.8%) | 34 (5.4%) |
|  | 4 indices out of tolerance | 33 (5.2%) | 3 (0.5%) | 21 (3.3%) |
|  | 5 indices out of tolerance | 18 (2.9%) | 0 (0%) | 16 (2.5%) |
|  | 6 indices out of tolerance | 9 (1.4%) | 0 (0%) | 9 (1.4%) |
|  | 7 indices out of tolerance | 2 (0.3%) | 0 (0%) | 0 (0%) |
|  | Total | 630 (100%) | 630 (100%) | 630 (100%) |

*3.2 Solution validation*

The distribution of the QA metric dose-volume indices for the data from 6 additional patients is shown in Figure 4. The 96% fractions for Bladder ΔD50 (%), the 100% fractions for FH-L ΔD25 (%) and the 96% fractions for FH-L ΔD40 (%) are within tolerance. However, only 54% of the fractions for PTV ΔCI (%) are within tolerance.

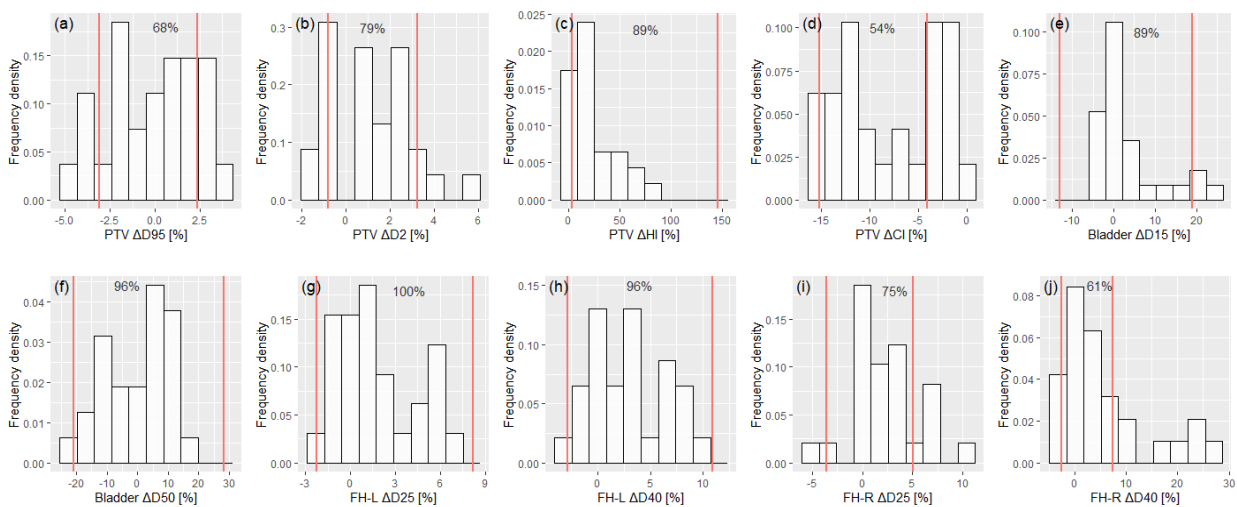

Figure 4. Distributions of the QA metric metrics of 28 validation fractions. Red vertical lines indicate the thresholds determined in Section 3.1.

Table 5 shows the number of dose-volume indices that are out of tolerance for 28 validation fractions. There were 5 fractions with four or more out-of-tolerance metrics in total deviation. After review of a physician, these 5 fractions are considered clinically acceptable.

Table 5. The number of out-of-tolerance dose-volume indices for validation fractions

| Number of out-of-tolerance indices for one fraction [indices] | Number of fractions | | |
|---|---|---|---|
| | Anatomical + delivery [fractions] | Anatomical [fractions] | Delivery [fractions] |
| All indices within tolerance | 12 (42.9%) | 13 (46.4%) | 22 (78.6%) |
| 1 index out of tolerance | 6 (21.4%) | 7 (25.0%) | 6 (21.4%) |
| 2 indices out of tolerance | 4 (14.3%) | 4 (14.3%) | 0 (0%) |
| 3 indices out of tolerance | 1 (3.6%) | 4 (14.3%) | 0 (0%) |
| 4 indices out of tolerance | 4 (14.3%) | 0 (0%) | 0 (0%) |
| 5 indices out of tolerance | 1 (3.6%) | 0 (0%) | 0 (0%) |
| Total | 28 (100%) | 28 (100%) | 28 (100%) |

*3.3 An example of solution application*

By examining the data, we found that one patient's data had a fraction with 5 QA metric dose-volume indices out of tolerance, including PTV ΔD95 (%), PTV ΔD2 (%), Bladder ΔD15 (%), FH-R ΔD25 (%), and FH-R ΔD40 (%).

For source of error analysis, this fraction had 3 anatomically relevant indicators out of tolerance, Bladder ΔD15 (%), FH-R ΔD25 (%), FH-R ΔD40 (%), and 1 delivery-relevant indicator out of tolerance, PTV ΔD2 (%).

The planning ROIs and daily recontoured ROIs of the fraction for the patient are shown in Figure 5. A shift of the right femoral head and an extension of the PTV can be observed. Our solution can quantitatively identify the dosimetric impact of these anatomical variations. The DVHs of the planning and QA metric indices are shown in Figure 6. A decrease in the dose of PTV and an increase in the dose of the FH-R can be observed.

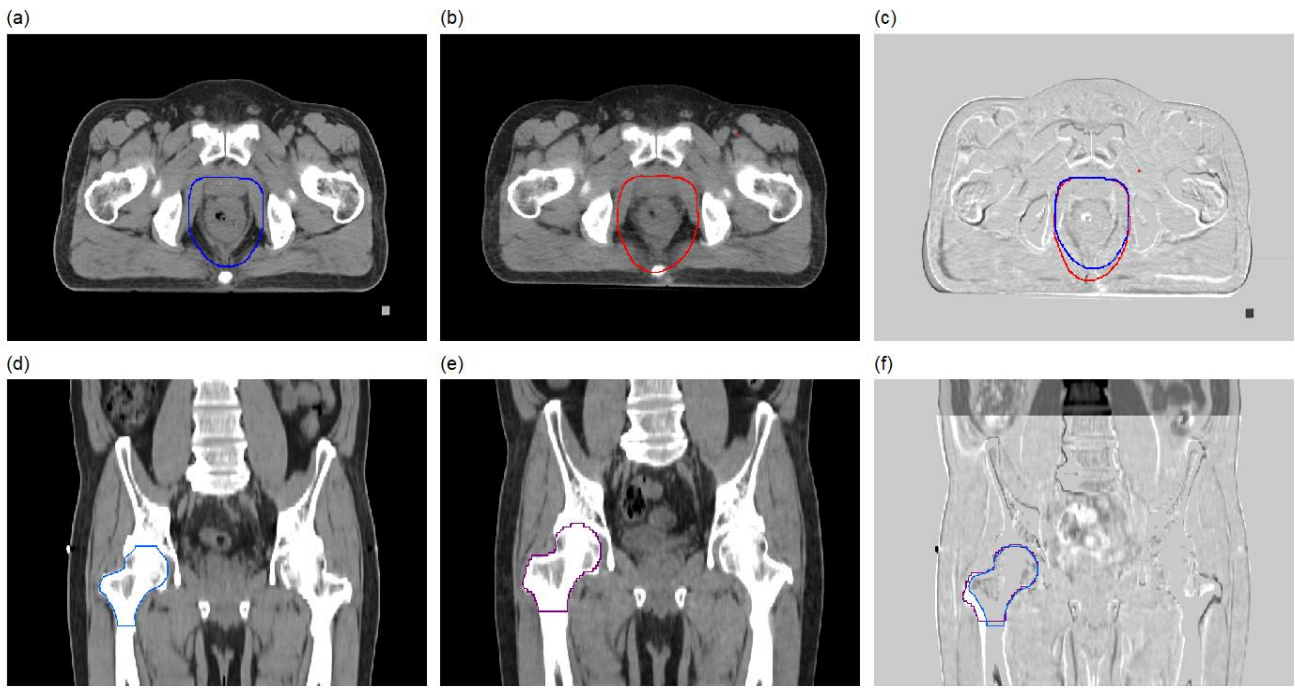

Figure 5. Comparison between the planning ROIs and daily recontoured ROIs after rigid registration. (a) The PTV on the planning CT scan; (b) The recontoured PTV on the daily CT scan; (c) The daily CT scan overlaid on the planning CT scan. The red delineation is the planning PTV, and the blue delineation is the recontoured PTV; (d) The right femoral head on the planning CT scan; (e) The recontoured right femoral head on the daily CT scan; and (f) The daily CT scan overlaid on the planning CT scan. The blue delineation is the planning right femoral head, and the purple delineation is the recontoured right femoral head. A shift of the right femoral head and an extension of the PTV can be obviously observed.

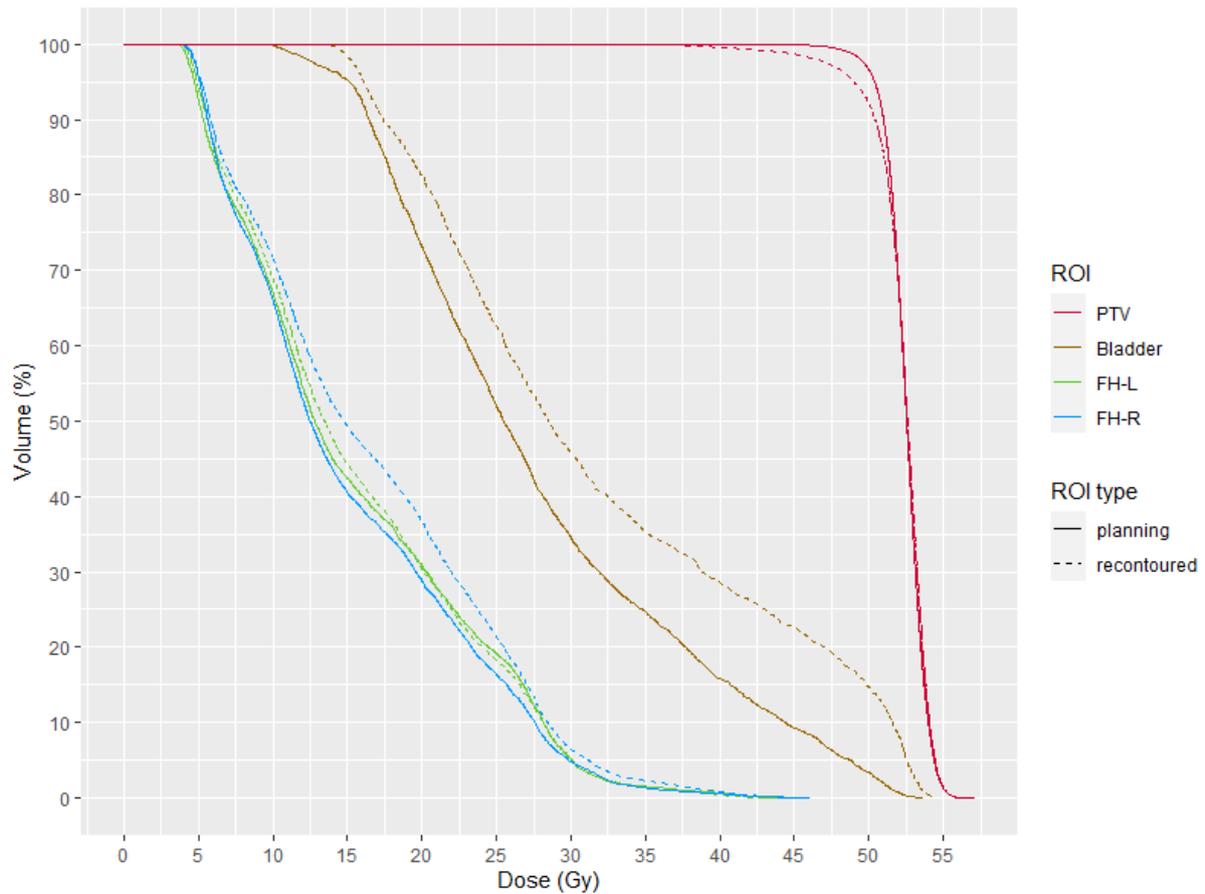

Figure 6. The DVHs of the planning and QA metric indices for the patient described above. The red curve is the PTV, the brown curve is the bladder, the green curve is the left femoral head and the blue curve is the right femoral head. The solid curve represents the ROIs used for the treatment plan on planning CT scans, the dashed curve represents the ROI that was recontoured on daily CT scans. The curve of the recontoured PTV decreased and the curve of the recontoured right femoral head increased compared to the original planning curve.

## 4. Discussion:

There are three major innovations in this study. First, the whole solution can directly connect to the dose volume indices that physicians and physicists are concerned about, and the impact on each ROI can be quantified. Second, by using 95% intervals, we proposed a setting method for the tolerance range based on the treatment history. Third, by calculating anatomically relevant and delivery relevant indices, more information could be provided to physicists for follow-up actions.

In this study, 45 fractions of 10 patients were enrolled for the tolerance setup, and 28 fractions of 6 patients were enrolled for validation, this is because recontouring ROIs is a time-consuming and labor-intensive task for physicians. For clinical use in future, more data and a more sophisticated workflow may be required for data process.

Since MR-Linac was introduced for clinical use, an increasing number of studies investigated how to perform online QA for ART plans.[44] Wang et al.[45] and Nachbar et al.[46] proposed a method to calculate dose distribution based on pretreatment MR images. However, without PTV recontouring and EPID dose reconstruction, this forward calculated dose distribution may still differ from the dose distribution patient received. For example, we forward calculated PTV D95 using daily CT scans with transferred ROIs after rigid registration based on Monte Carlo algorithm, the Pearson correlation coefficient between this type of PTV ΔD95 (%) and the QA metric for the 45 patients is moderate ($\rho$=0.573).

By using EPID reconstructed 3D dose distribution, Olaciregui-Ruiz et al.[41,47] proposed an automatic dosimetric verification method that can be implemented on MR-Linac. Two comparisons were performed, including gamma analysis (3%/2 mm/10% threshold) and comparison of the median dose to the high-dose volume (HDV ΔD50). Eighty-five percent for the gamma pass rate and 5% for HDV ΔD50 were used as tolerance limit values. These two indices may not relate to the indices that are of concern to the physician. Further sensitivity studies are necessary for the optimal determination of indicators and tolerance limit values.[41,47]

In error source analysis, for nonerror-introduced fractions, all delivery-relevant indices were within tolerance, and the distribution of the anatomically relevant indices was similar to the distribution of total deviation (Table 4). For error introduced fractions, the distribution of the anatomically relevant indices was similar to the nonerror-introduced fractions (Table 4). However, the number of delivery relevant indices appeared. For the fractions with more than 4 out-of-tolerance indices, the distribution of delivery relevant indices was similar to the distribution of the total deviation. This indicated that our solution has some potential for distinguishing error sources.

During solution validation, the values of two indices [PTV ΔCI (%) and FH-R ΔD40 (%)] had relatively low proportions within the tolerance. After reviewing the data, we found that two patients had all fractions out of tolerance in these two indices (the patient whose data is shown in Figures 5 and 6 is one of them). These two patients may have a deviation in setup between planning and treatment.

This study has some limitations. In the tolerance setting, 95% interval and 4 out-of-tolerance QA metrics were partly subjective based. The main reason is that there is not enough data. As we mentioned above, recontouring ROIs is a time-consuming and labor-intensive task for physicians. With accumulating more data, these setting can be improved. Second, anatomical variances during treatment, such as the movement of the patient's organs, were not considered in this study.

Our proposed method is an ART QA solution. Because the indices used in this solution were clinically relevant dose-volume indices, this solution can also be used for ART patient monitoring, especially for anatomically relevant indices. Further data and research are required in this direction.

**Conclusion:**

This aim of this article was to develop a clinically relevant online patient quality assurance (QA) solution, and the online patient QA solution proposed using daily CT scans and EPID-based in vivo dosimetry is clinically feasible. Tolerance and action level setting using distribution of dose-volume index is comparatively reasonable. Source of error analysis has the potential for distinguishing sources of error and guiding ART for future treatments. A research based on more data for clinical use may be investigated in future.

**Acknowledgments:**

This work is supported by the Shanghai Committee of Science and Technology Fund (19DZ1930902, 21Y21900200), Xuhui District Artificial Intelligence Medical Hospital Cooperation Project (2021-012) and National Natural Science Foundation of China, Grant/Award Numbers: 11675042.

**Conflicts of Interest:**

Y. L. is employed by United Imaging Healthcare Corporation.

**Data availability statement:**

Authors will share data upon request to the corresponding author.